\documentclass[aps,prd,preprintnumbers,superscriptaddress,nofootinbib,twocolumn]{revtex4-1}
\usepackage[dvipdfmx]{graphicx}
\usepackage{bm,latexsym,amsmath,amssymb,amsfonts,mathrsfs}
\usepackage{color}
\input{colordvi.tex}
\allowdisplaybreaks[1]
\usepackage{url}
\hypersetup{
    colorlinks=true,
    citecolor=cyan,
    linkcolor=magenta,
}
\newcommand*{\D}{\textrm{d}}
\newcommand*{\mpl}{M_\textrm{Pl}}
\definecolor{pn}{rgb}{1.0,0.2,0.8}

\begin{document}
\title{Revisiting slow-roll dynamics and the tensor tilt in general single-field inflation}
\author{Yosuke~Mishima}
\email[Email: ]{yosuke"at"rikkyo.ac.jp}
\affiliation{Department of Physics, Rikkyo University, Toshima, Tokyo 171-8501, Japan
}
\author{Tsutomu~Kobayashi}
\email[Email: ]{tsutomu"at"rikkyo.ac.jp}
\affiliation{Department of Physics, Rikkyo University, Toshima, Tokyo 171-8501, Japan
}
\begin{abstract}
We explore the possibility of
a blue-tilted gravitational wave spectrum
from potential-driven slow-roll inflation in the Horndeski theory.
In Kamada \textit{et al.} (2012), it was claimed that a blue gravitational wave spectrum
cannot be obtained
from stable potential-driven slow-roll inflation within the Horndeski framework.
However, it has been demonstrated that the spectrum of primordial gravitational
waves can be blue in inflation with the Gauss-Bonnet term,
where the potential term is dominant and
slow-roll conditions as well as the stability conditions are satisfied.
To fill in this gap,
we clarify where the discrepancy is coming from.
We extend the formulation of Kamada \textit{et al.} (2012) and
show that a blue gravitational wave spectrum
can certainly be generated
from stable slow-roll inflation if some of the
conditions previously imposed on the form of the free functions
in the Lagrangian are relaxed.
\end{abstract}
\preprint{RUP-19-30}
\maketitle
\section{Introduction}

Inflation~\cite{Guth:1980zm,Starobinsky:1980te,Sato:1980yn} is
the most promising candidate of the scenario for the early Universe,
explaining naturally the large-scale homogeneity of the
observed Universe and the origin of primordial density perturbations
that lead to the CMB fluctuations and the large-scale structure.
Inflation also predicts the existence of
a stochastic background of primordial
gravitational waves (tensor perturbations),
which has yet to be observed.
In the most standard inflationary scenario,
a quasi-de Sitter expansion is driven by
the potential of a slowly rolling canonical scalar field.
A robust prediction of this standard scenario is
that the tensor spectral tilt, $n_t$, is
always negative, $n_t=-r/8<0$, where $r$ is the tensor-to-scalar ratio.
Any signature of a blue-tilted ($n_t>0$) gravitational
wave spectrum therefore
implies a nonstandard model of inflation
(or even an alternative scenario to inflation).

Going beyond the standard slow-roll inflation models,
a number of models with a modified structure of
inflaton's kinetic term and/or
nonminimal coupling between gravity and inflaton
have been proposed.
Such diverse inflation models can be
described in a unified manner
by the generalized G-inflation framework~\cite{Kobayashi:2011nu},
which is based on the Horndeski
theory/generalized Galileons~\cite{Horndeski:1974wa,Deffayet:2011gz},
i.e., the most general scalar-tensor theory with second-order
field equations (see~\cite{Kobayashi:2019hrl} for a review).
Focusing on the (rather conservative)
case of \emph{potential-driven slow-roll inflation},
a generic way of describing such inflation models
within the Horndeski theory has been developed in~\cite{Kobayashi:2011nu,Kamada:2010qe,Kamada:2012se}.
While a blue-tilted gravitational wave spectrum
can indeed
be obtained in \emph{kinetically driven} G-inflation
thanks to stable violation of the null energy condition~\cite{Kobayashi:2010cm}
(see also~\cite{Cai:2014uka, Cai:2015yza, Cai:2016ldn}),
it is argued in~\cite{Kamada:2012se} that
$n_t>0$ can never be possible
as long as inflation is driven by a potential of
a slowly rolling scalar field.
\footnote{See, e.g.,
Refs.~\cite{Gruzinov:2004ty,Endlich:2012pz,Cannone:2014uqa,Bartolo:2015qvr,%
Ricciardone:2016lym,Fujita:2018ehq,Ashoorioon:2014nta}
for more radical inflation models
with a blue-tilted gravitational wave spectrum.}

While the argument in~\cite{Kamada:2012se} seems to be
general to a large extent, a counterexample is known to exist
in the literature: inflation with the nonminimal coupling to
the Gauss-Bonnet term~\cite{Satoh:2008ck,Satoh:2010ep,Guo:2010jr,%
Jiang:2013gza,Koh:2014bka,Koh:2016abf,Bhattacharjee:2016ohe,vandeBruck:2016xvt,%
Nozari:2017rta,Wu:2017joj,Koh:2018qcy,Chakraborty:2018scm}.
Though in this case the energy density of the slowly rolling inflaton
is dominated by its potential,
one can have a positive tensor tilt~\cite{Satoh:2008ck,Satoh:2010ep,Koh:2014bka}.
Since the nonminimal coupling between a scalar field and the Gauss-Bonnet term
is just a specific example of the Horndeski Lagrangian~\cite{Kobayashi:2011nu},
there must be something overlooked in the analysis of~\cite{Kamada:2012se}.

The purpose of this paper is to fill in the gap
between the above apparently contradicting statements.
In fact, the formulation of~\cite{Kamada:2012se}
is not general enough to accommodate
Gauss-Bonnet inflation. Moreover, an unnecessarily strong assumption
was made in~\cite{Kamada:2012se}. In this paper,
we improve these points and enlarge a possible model space of
slow-roll inflation within the Horndeski theory,
showing that blue gravitational waves can indeed be generated
from (stable) slow-roll inflation.

This paper is organized as follows.
In the next section,
we review the previous study~\cite{Kamada:2012se}
and suggest a possible improvement
as implied by the example of Gauss-Bonnet inflation.
In Sec.~\ref{Section:III},
we extend the slow-roll dynamics
to cover the inflationary model space
which has not been explored in~\cite{Kamada:2012se}.
We then discuss the possibility of a blue-tilted gravitational
wave spectrum from potential-driven slow-roll inflation
in Sec.~\ref{Section:IV}.
Finally, we draw our conclusion in Sec.~\ref{Section:V}.

\section{Slow-roll inflation from Horndeski}
\label{Section:II}

\subsection{A quick recap of Kamada \textit{et al.}~\cite{Kamada:2012se}}
\label{Section:II-1}

Let us review briefly the argument of Ref.~\cite{Kamada:2012se},
where generic slow-roll inflationary dynamics is investigated.
The analysis of Ref.~\cite{Kamada:2012se} is based on the
Horndeski theory, i.e., the most general scalar-tensor theory
with second-order field equations, whose Lagrangian is given by
\begin{align}
{\cal L}&=G_2(\phi, X) -G_3(\phi,X)\square\phi+ G_4(\phi,X)R
\notag \\ &\quad
 +\frac{\partial G_{4}}{\partial X}\left[(\square\phi)^2-(\nabla_\mu\nabla_\nu\phi)^2\right]
+G_5(\phi,X)G_{\mu\nu}\nabla^\mu\nabla^\nu\phi
\notag \\ & \quad
-\frac{1}{6}\frac{\partial G_{5}}{\partial X}\left[(\square\phi)^3
-3(\square\phi)(\nabla_\mu\nabla_\nu\phi)^2
+2(\nabla_\mu\nabla_\nu\phi)^3\right].
\end{align}
Here, $\phi$ is the scalar field, $X:= -g^{\mu\nu}\nabla_\mu\phi\nabla_\nu\phi/2$,
$R$ is the Ricci scalar, and $G_{\mu\nu}$ is the Einstein tensor.
The background cosmological equations and
the quadratic action governing cosmological perturbations
in the Horndeski theory
are found in Ref.~\cite{Kobayashi:2011nu}.

To describe the slow-roll dynamics of the scalar field during generic
potential-driven inflation,
it is assumed in Ref.~\cite{Kamada:2012se} that
the functions in the Lagrangian can be expanded in terms of $X$ as
\begin{align}
G_a=g_a(\phi)+h_a(\phi) X + {\cal O}(X^2)\quad (a=2,3,4,5).\label{Kamada-assumption-1}
\end{align}
The Taylor-expanded form~\eqref{Kamada-assumption-1} is
one of the central assumptions made in Ref.~\cite{Kamada:2012se}.
Since $g_2$ corresponds to the potential, hereafter we will write $g_2=-V(\phi)$.
$g_4$ plays the role of the effective Planck mass squared.
We assume that $g_4>0$, which we will confirm
is equivalent to the stability condition for tensor perturbations.
Note that $g_3$ and $g_5$ can be absorbed into
the redefinition of the other functions of $\phi$ with the help of
integration by parts,
and hence we may set $g_3=0=g_5$ without loss of generality~\cite{Kobayashi:2011nu}.
Therefore, we essentially have six functions of $\phi$
(including the potential $V$)
characterizing general slow-roll inflation driven by the potential.
By taking these functions appropriately one can, for example, describe
the known variants of Higgs inflation
such as nonminimal Higgs inflation~\cite{Bezrukov:2007ep},
new Higgs inflation~\cite{Germani:2010gm}, and Higgs G-inflation~\cite{Kamada:2010qe}.

During slow-roll inflation,
we may assume that the time variation of $g_a(\phi)$ and $h_a(\phi)$
is small. We are thus led to the following slow-roll conditions:
\begin{align}
&\epsilon:=-\frac{\dot H}{H^2}\ll 1,
\quad
\eta := -\frac{\ddot\phi}{H\dot\phi}\ll 1,
\notag \\ &
\delta_M:=\frac{\dot g_4}{Hg_4}\ll 1,
\quad
\frac{\dot h_a}{Hh_a}\ll 1,\label{slow-roll-conditions}
\end{align}
where $H$ is the Hubble parameter and a dot
denotes differentiation with respect to the cosmic time.
With some manipulation, the background cosmological
equations under these conditions reduce to~\cite{Kamada:2012se}
\begin{align}
6g_4H^2&\simeq V,\label{kmdeq1}
\\
-4g_4\dot H+2\dot g_4 H&\simeq
\dot\phi^2\left(
u+3vH\dot\phi
\right),\label{kmdeq2}
\\
3H\dot \phi &\simeq
\frac{1}{2v}\left(-u+\sqrt{u^2-4U'v}\right),\label{kmdeq3}
\end{align}
where we defined
\begin{align}
&
u(\phi):=h_2+\frac{h_4V}{g_4},
\quad
v(\phi):=h_3+\frac{h_5 V}{6g_4},
\notag \\ &
U'(\phi):=g^2_4\left(\frac{V}{g^2_4}\right)',\label{defuvU}
\end{align}
and a prime denotes differentiation with respect to $\phi$.
It is obvious that Eq.~\eqref{kmdeq1} is essentially the
Friedmann equation.
Equations~\eqref{kmdeq2} and~\eqref{kmdeq3}
correspond respectively to the familiar equations
$-2\mpl^2 \dot H = \dot\phi^2$
and
$3 H\dot \phi\simeq -V'$ in the canonical slow-roll inflation model.

In Ref.~\cite{Kamada:2012se} it is assumed that
\begin{align}
u(\phi) > 0, \label{kmd-assume}
\end{align}
probably because $u$ determines the sign of the kinetic term
of $\phi$ in the simple case with $h_4=0$
and hence is expected to be correlated with
some of the stability conditions.
This is another assumption
which we revisit carefully in this paper.

Using Eqs.~\eqref{kmdeq2} and~\eqref{kmdeq3} we obtain
\begin{align}
2\epsilon+\delta_M=\frac{\dot\phi^2}{4g_4H^2}
\left(u+\sqrt{u^2-4U'v}\right).\label{2de000}
\end{align}
It follows from the assumption~\eqref{kmd-assume} that
\begin{align}
2\epsilon + \delta_M >0.\label{2ed0}
\end{align}
This inequality will lead to the important conclusion
on the tensor tilt.

The quadratic action for tensor perturbations
in generic potential-driven inflation
is given by~\cite{Kamada:2012se}
\begin{align}
S^{(2)}_h=\frac{1}{4}\int\D t\D^3x\, a^3 g_4
\left[
\dot h_{ij}^2 - a^{-2}(\partial_kh_{ij})^2
\right].\label{tensor-action}
\end{align}
A nonstandard feature appears only in the
time-dependent effective Planck mass $g_4$.
The stability condition is equivalent to
the aforementioned assumption $g_4>0$. Following the
usual quantization procedure one obtains the tensor power spectrum,
\begin{align}
{\cal P}_h = \frac{H^2}{\pi^2 g_4},\label{eq:power_in_J}
\end{align}
and its spectral index,
\begin{align}
n_t=-2\epsilon-\delta_M.
\end{align}
Thus, from Eq.~\eqref{2ed0} we see that the tensor power spectrum
would never be blue
in generic potential-driven inflation,
\begin{align}
n_t<0.
\end{align}

The quadratic action for the curvature perturbation
in the unitary gauge is given by~\cite{Kamada:2012se}
\begin{align}
S_\zeta^{(2)}=\int \D t\D^3x\, a^3\frac{{\cal F}}{c_s^2}
\left[\dot\zeta^2-a^{-2}c_s^2(\partial\zeta)^2\right],
\label{scalar-action}
\end{align}
where
\begin{align}
{\cal F}&:= \frac{\dot\phi^2}{2H^2}\left(u +4v H\dot\phi \right),
\\
c_s^2&:=\frac{u+4vH\dot\phi }{u+6vH\dot\phi }.
\end{align}
Using Eq.~\eqref{kmdeq3} we obtain
\begin{align}
{\cal F}&=\frac{\dot\phi^2}{6H^2}\left(u+2\sqrt{u^2-4U'v}\right),
\\
c_s^2&=\frac{1}{3}\left(2+\frac{u}{\sqrt{u^2-4U'v}}\right).
\end{align}
Thus, under the assumption~\eqref{kmd-assume},
the stability conditions ${\cal F}>0$ and $c_s^2>0$
are indeed satisfied.
However, $u>0$ seems to be only a \emph{sufficient} condition for the stability.

It is straightforward to calculate the power spectrum
and the spectral index~\cite{Kobayashi:2011nu}:
\begin{align}
{\cal P}_\zeta& =\frac{1}{8\pi^2}\frac{H^2}{c_s{\cal F}},
\\
n_s-1&=-2\epsilon-\frac{\dot c_s}{Hc_s}-\frac{\dot {\cal F}}{H{\cal F}}.
\label{nsformula}
\end{align}

\subsection{Possible improvement of Kamada \textit{et al.}
~\cite{Kamada:2012se}: %
The case of Gauss-Bonnet inflation}
\label{Section:II-2}

It was pointed out that the tensor spectral index can be positive
in (stable) slow-roll inflation with
the Gauss-Bonnet term~\cite{Satoh:2008ck,Satoh:2010ep,Koh:2018qcy}.
More explicitly, in Refs.~\cite{Satoh:2008ck,Satoh:2010ep,Koh:2018qcy}
the following nonminimal coupling to the Gauss-Bonnet term is considered:
\begin{align}
f(\phi)\left(R^2-4R_{\mu\nu}R^{\mu\nu}+R_{\mu\nu\rho\sigma}R^{\mu\nu\rho\sigma}\right).
\label{GB-term}
\end{align}
It is well known that this term yields second-order field equations,
and hence resides within the Horndeski theory.
Nevertheless, the result of Refs.~\cite{Satoh:2008ck,Satoh:2010ep,Koh:2018qcy}
seems to be inconsistent with the conclusion of Ref.~\cite{Kamada:2012se}.
This indicates that the work of~\cite{Kamada:2012se},
which was probably built upon somewhat
stronger assumptions than necessary,
can potentially be improved to accommodate
potential-driven inflation models with a blue tensor spectrum such as
Gauss-Bonnet inflation.

Let us start with looking at
how the nonminimal coupling to the Gauss-Bonnet term~\eqref{GB-term}
is incorporated into the Horndeski theory.
As shown in Ref.~\cite{Kobayashi:2011nu},
the way of reproducing the term~\eqref{GB-term}
from the Horndeski functions is nontrivial:
\begin{align}
  &G_2 \supset 8f'''' X^2(3-\ln X), \quad
  G_3 \supset 4f''' X(7 -3\ln X), \notag\\
  &G_4 \supset 4f''X(2-\ln X), \quad
  G_5 \supset -4f'\ln X. \label{GBtermfromHorndeski}
\end{align}
This clearly shows that the Taylor-expanded form~\eqref{Kamada-assumption-1}
fails to capture the structure of the Gauss-Bonnet term.
The first three terms could be slow-roll suppressed
because they are proportional to second or higher derivatives
of weakly $\phi$-dependent function $f$,
but $G_5 \supset -4f'\ln X$ cannot be ignored even in the slow-roll regime.
This observation
hints at how we can proceed to extend the framework of~\cite{Kamada:2012se}.

Moreover, we have seen that the previous assumption~\eqref{kmd-assume}
is likely to be too strong. It is therefore desirable to revisit this point and
clarify the \emph{necessary} condition for the stability.
This will also enlarge the possible model space of slow-roll inflation
explored by Ref.~\cite{Kamada:2012se}.

\section{Slow-roll dynamics}
\label{Section:III}

To capture the essential part of the Gauss-Bonnet term
in the slow-roll regime,
let us now assume that the Horndeski functions take the form of
\begin{align}
G_a=g_a(\phi)+\lambda_a(\phi)\ln X +h_a(\phi)X+ \cdots,
\label{lnx-expansion}
\end{align}
where the ellipsis stands for slow-roll suppressed terms.
As in the previous analysis, we eliminate $g_3$ and $g_5$
by performing integration by parts, and assume that $g_4>0$.
One may also consider the terms of the form $\xi_a(\phi)X\ln X$,
which could be as large as, or even larger than, $h_a X$.
In the present analysis, however, we will assume that
$\lambda_a$ is already of first order in the slow-roll approximation,
$\lambda_a={\cal O}(h_aX)$, and
accordingly $\xi_a X\ln X$ is of second order.
The newly introduced terms $\lambda_a \ln X$ would be dangerous
in the $X\to 0$ limit. However, as we will see below, at least some of them
yield only regular terms at the level of field equations.

We assume the same slow-roll conditions as given
in Eq.~\eqref{slow-roll-conditions}. For the new functions
we impose the analogous slow-roll conditions,
\begin{align}
\frac{\dot\lambda_a}{H\lambda_a}\ll1 .
\end{align}
Under these assumptions, the time-time component of
the gravitational field equations
for a cosmological background
reduces to
\begin{align}
6g_4H^2&\simeq V+\lambda_2(2-\ln X) -6\lambda_4 H^2\ln X
\notag \\ &\quad
+6\lambda_3H\dot\phi+6\lambda_5H^3\dot\phi .\label{Frlambdalambda}
\end{align}
To avoid the singular terms in the $X\to 0$ limit, we require that
\begin{align}
\lambda_2=0,\quad \lambda_4=0.
\end{align}
As seen from Eq.~\eqref{Frlambdalambda},
$\lambda_3$ and $\lambda_5$ do not lead to any
singular terms in the field equations, and hence are acceptable.
In addition to the slow-roll conditions, we impose the following
\emph{potential-dominance} conditions on these two functions:
\begin{align}
  \delta_3:=\frac{\lambda_3 \dot\phi}{g_4H}
  \lesssim {\cal O}(\epsilon), \quad
\delta_5:=\frac{\lambda_5H\dot\phi}{g_4} \lesssim {\cal O}(\epsilon),
\end{align}
namely, the last two terms on the right-hand side of
Eq.~\eqref{Frlambdalambda} are actually of the same order of
the other slow-roll suppressed terms.
We thus have the same equation as Eq.~\eqref{kmdeq1},
\begin{align}
6g_4 H^2\simeq V,\label{hamcons}
\end{align}
even in the presence of the $\ln X$ terms.
This equation rules out the possibility of $g_4<0$ (for $V>0$).

The space-space components of the gravitational field equations
and the equation of motion for the scalar field
in the slow-roll regime reduce to
\begin{align}
-4g_4\dot H+2\dot g_4H &\simeq \dot\phi^2{\cal I},\label{eveq}
\\
3H\dot\phi{\cal I}&\simeq  -U'(\phi),\label{sceq1}
\end{align}
where
\begin{align}
{\cal I}&:=u(\phi)+3v(\phi)H\dot\phi +\frac{\varpi(\phi)}{3H\dot\phi},
\\
\varpi(\phi)&:=\varpi_3(\phi)+\varpi_5(\phi),
\\
\varpi_3(\phi)&:=\frac{3V}{g_4}\lambda_3,
\quad
\varpi_5(\phi):=\frac{1}{6}\left(\frac{V}{g_4}\right)^2\lambda_5 ,\label{defvarpi3and5}
\end{align}
and $u$, $v$, and $U'$ were defined earlier in Eq.~\eqref{defuvU}.
Now it is easy to solve Eq.~\eqref{sceq1} for $3H\dot\phi$ to get
\begin{align}
3H\dot\phi \simeq \frac{1}{2v}\left[
-u\pm \sqrt{u^2-4v(U'+\varpi)}
\right].\label{3hdot}
\end{align}
At this stage we have two branches (if $v\neq 0$),
but it will turn out in the end that
the ``$-$'' branch exhibits instabilities
of scalar perturbations.
Therefore, here we only consider the ``$+$'' branch.
Using Eq.~\eqref{3hdot}, one can rewrite ${\cal I}$
in terms of the functions of $\phi$ as
\begin{align}
{\cal I}=\frac{U'}{2(U'+\varpi)}\left[u+\sqrt{u^2-4v(U'+\varpi)}\right].
\end{align}
If $v=0$, we do not need to care about the branches and we instead have
\begin{align}
3H\dot\phi &\simeq -\frac{U'+\varpi}{u},
\\
{\cal I}&=\frac{U' u}{U'+\varpi}.\label{iv0}
\end{align}

Equation~\eqref{eveq} can also be expressed as
\begin{align}
2\epsilon+\delta_M = \frac{\dot\phi^2{\cal I}}{2g_4H^2}.\label{2deIII}
\end{align}
This is the generalization of Eq.~\eqref{2de000}
and will be used later in the next section.

Let us take a look at the role of ${\cal I}$
in the slow-roll dynamics, focusing on the simple case with $g_4=\mpl^2/2$.
Using the background equations~\eqref{hamcons},~\eqref{eveq},
and~\eqref{sceq1},
the potential slow-roll parameter, $\epsilon_V:=\mpl^2(V'/V)^2/2$,
can be written as
\begin{align}
\epsilon =\frac{\epsilon_V}{{\cal I}}.
\end{align}
This implies that, even if the potential is
too steep to support usual inflation (say, $\epsilon_V={\cal O}(1)$),
inflation can still occur provided that $|{\cal I}|\gg 1$.

To highlight the impact of the newly introduced term,
let us further focus on the case with $u=1$ and $v=0$.
In this case, the Lagrangian is given by
${\cal L}=(\mpl^2/2)R+X-V+\;$($\ln X\;$corrections).
We then have
\begin{align}
&{\cal I}=\frac{V'}{V'+\varpi}
\;\;\Rightarrow\;\;\epsilon = \frac{\mpl^2}{2}\frac{V'(V'+\varpi)}{V^2},
\\
&3H\dot\phi \simeq -(V'+\varpi).
\end{align}
Thus, $\varpi$ effectively shifts the potential slope.
If the potential is nearly flat and $\varpi\lesssim{\cal O}(V')$,
the $\ln X$ terms have only a minor effect on the dynamics.
In the opposite limit, $\varpi \gg {\cal O}(V')$, inflation is spoiled.
The most interesting case is that $V'$ could be large but
is canceled by $\varpi$: $V'\simeq - \varpi$.
In this case, inflation occurs
with the help of the $\ln X$ terms.

It should be emphasized that
so far we have made no assumption about the sign of $u$.
Only the constraint
coming from the background dynamics is that
the expression in the square root must be non-negative:
\begin{align}
u^2\ge 4v(U'+\varpi).\label{rootplus}
\end{align}

\section{Tensor tilt and stability}
\label{Section:IV}

Let us move to the main question of this paper:
\emph{Is a blue tensor spectrum compatible with the
potential-driven slow-roll dynamics and the stability of
cosmological perturbations?}

We substitute the assumed form of
the Horndeski functions $G_a$ [Eq.~\eqref{lnx-expansion}]
to the general formulas of the quadratic action for
cosmological perturbations derived in Ref.~\cite{Kobayashi:2011nu}.
We then make the slow-roll and potential-dominance approximations.
Even if one takes into account the $\ln X$ terms in $G_a$,
it is found that under these approximations
the quadratic action for tensor perturbations
remains the same as Eq.~\eqref{tensor-action}.
Therefore, the tensor spectral index is given by
\begin{align}
n_t=-2\epsilon-\delta_M = - \frac{\dot\phi^2{\cal I}}{2g_4H^2},
\end{align}
where we used Eq.~\eqref{2deIII}.
Thus, the sign of ${\cal I}$ plays the key role in
determining the sign of $n_t$.

The quadratic action for the curvature perturbation
takes the form of Eq.~\eqref{scalar-action}, but now with
\begin{align}
{\cal F}&=\frac{\dot\phi^2}{2H^2}\left(u+
4vH\dot\phi +\frac{4\varpi_3}{9H\dot\phi}\right),
\\
c_s^2&=
\frac{u+4vH\dot\phi + 4\varpi_3/9H\dot\phi}{u+6vH\dot\phi}.
\end{align}
For the stability of the scalar sector it is necessary
that both denominator and numerator of $c_s^2$ are positive.
Using Eq.~\eqref{3hdot}, we have
\begin{align}
u+6vH\dot\phi =
\begin{cases}
    \pm \sqrt{u^2-4v(U'+\varpi)} \;\;&(v\neq 0)\\
    u &(v=0)
  \end{cases}.
\end{align}
Thus, as long as we take the ``$+$'' branch,
one of the stability conditions is automatically satisfied.
However, in the case of $v=0$, the stability condition requires $u>0$.
The numerator of $c_s^2$ reads
\begin{align}
&u+4vH\dot\phi + 4\varpi_3/9H\dot\phi
\notag \\ &=
\begin{cases}
  \frac{(1-2A)}{3}u+\frac{2(1-A)}{3}\sqrt{u^2-4v(U'+\varpi)} &(v\neq 0) \\
  (1-\frac{4A}3)u &(v=0)
\end{cases},
\end{align}
where $A:=\varpi_3/(U'+\varpi)$.
Since several independent functions participate in
the stability conditions, it is not so illuminating
to analyze the most general case.
Instead let us consider the following two special cases:
(i) $\varpi_3=\varpi_5=0$, and (ii) $v=0$.

\subsection*{(i) $\varpi_3=\varpi_5=0$}
\label{Section:IV-1}

  \begin{figure}[tb]
    \begin{center}
            \includegraphics[keepaspectratio=true,height=70mm]{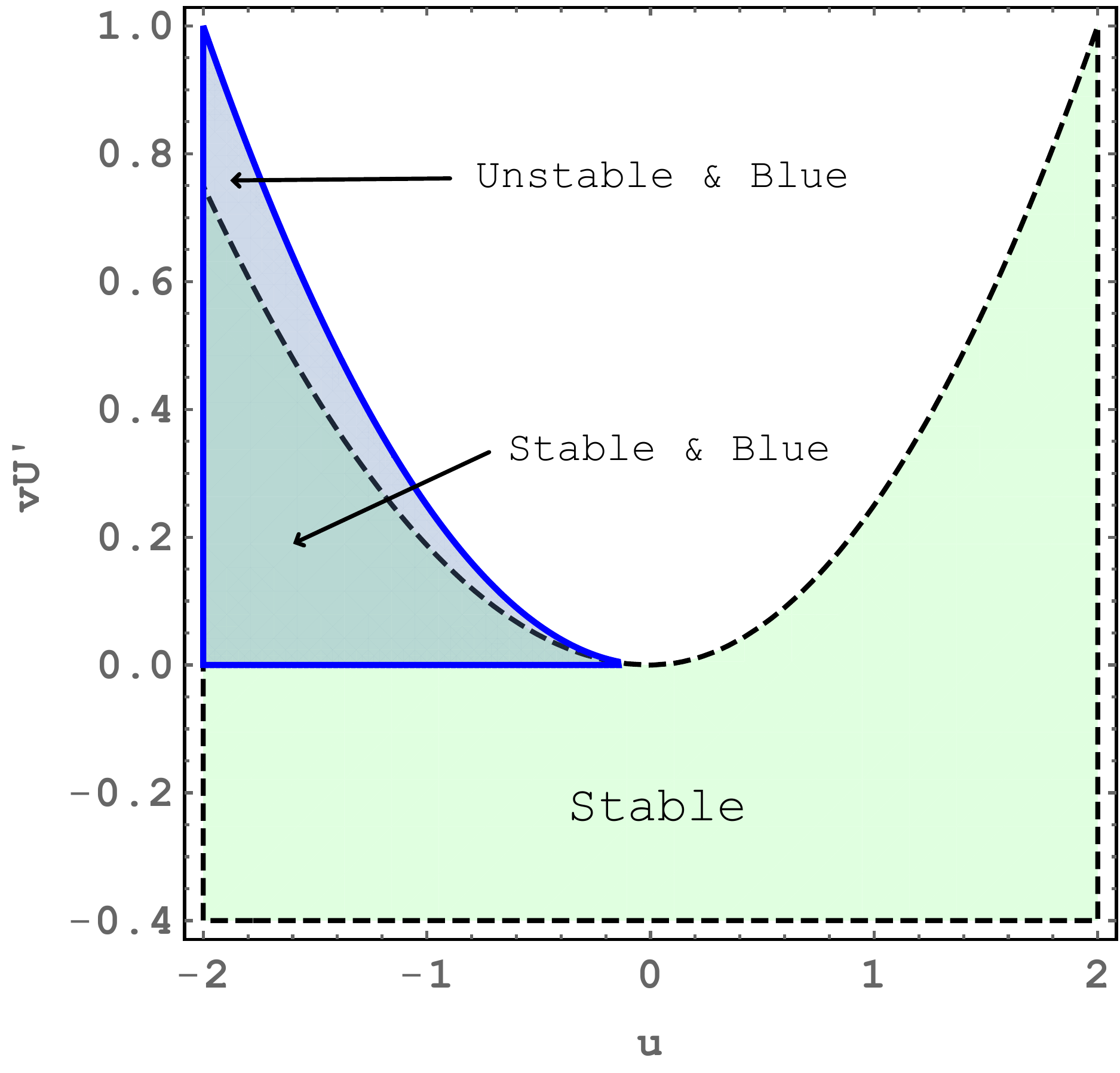}
    \end{center}
       \caption{Tensor tilt and stability in the reanalysis of~\cite{Kamada:2012se}.
       The $u<0$ region was overlooked in the previous analysis.
       Note in passing that the white region does not satisfy the
       condition~\eqref{rootplus}.
  	}
       \label{fig:re-kamada.pdf}
  \end{figure}

Since the $\ln X$ terms vanish in $G_a$,
this case corresponds to the reanalysis of Ref.~\cite{Kamada:2012se}.
The tensor tilt and the stability are determined by the
two combinations of the functions, $u$ and $vU'$.
The stability conditions read
\begin{align}
u+2\sqrt{u^2-4vU'}>0,
\end{align}
while the tensor tilt depends on the sign of
\begin{align}
{\cal I}=\frac{1}{2}\left(
u+\sqrt{u^2-4vU'}
\right).
\end{align}
We plot in Fig.~\ref{fig:re-kamada.pdf}
the stable region in the $(u,vU')$ plane.
It is found that the scalar perturbations are stable
and $n_t>0$ if
\begin{align}
u<0,\quad 0<vU'<\frac{3}{16}u^2.\label{standbl}
\end{align}
To realize $n_t>0$, it is \emph{not} necessary to extend
the formulation of~\cite{Kamada:2012se}.
The region~\eqref{standbl} was just overlooked in the previous analysis.

Let us present a simple explicit example
whose Lagrangian is given by
\begin{align}
{\cal L}=\frac{\mpl^2}{2}R-X-V(\phi)-\frac{\alpha}{\Lambda^3}X\Box\phi,
\end{align}
where the potential is approximated by
$V=\Lambda^3\phi$ in the regime we are interested in.
For $0<\alpha<3/16$ the above conditions are satisfied.
One obtains a slow-roll inflationary solution,
\begin{align}
H\simeq \frac{\Lambda^2}{(6\mpl^2|{\cal I}|)^{1/3}}t^{1/3},
\quad
\phi \simeq \frac{3\mpl^{2/3}\Lambda}{(6|{\cal I}|)^{2/3}}t^{2/3},
\end{align}
for $t\gg \mpl^{1/2}\Lambda^{-3/2}|{\cal I}|^{1/4}$,
where ${\cal I}=(-1+\sqrt{1-4\alpha})/2=\,$const.
Note that $H\gg \Lambda^{3/2}\mpl^{-1/2}|{\cal I}|^{-1/4}$ and hence
$\Lambda$ must be sufficiently low.

\subsection*{(ii) $v=0$}
\label{Section:IV-2}

  \begin{figure}[tb]
    \begin{center}
            \includegraphics[keepaspectratio=true,height=70mm]{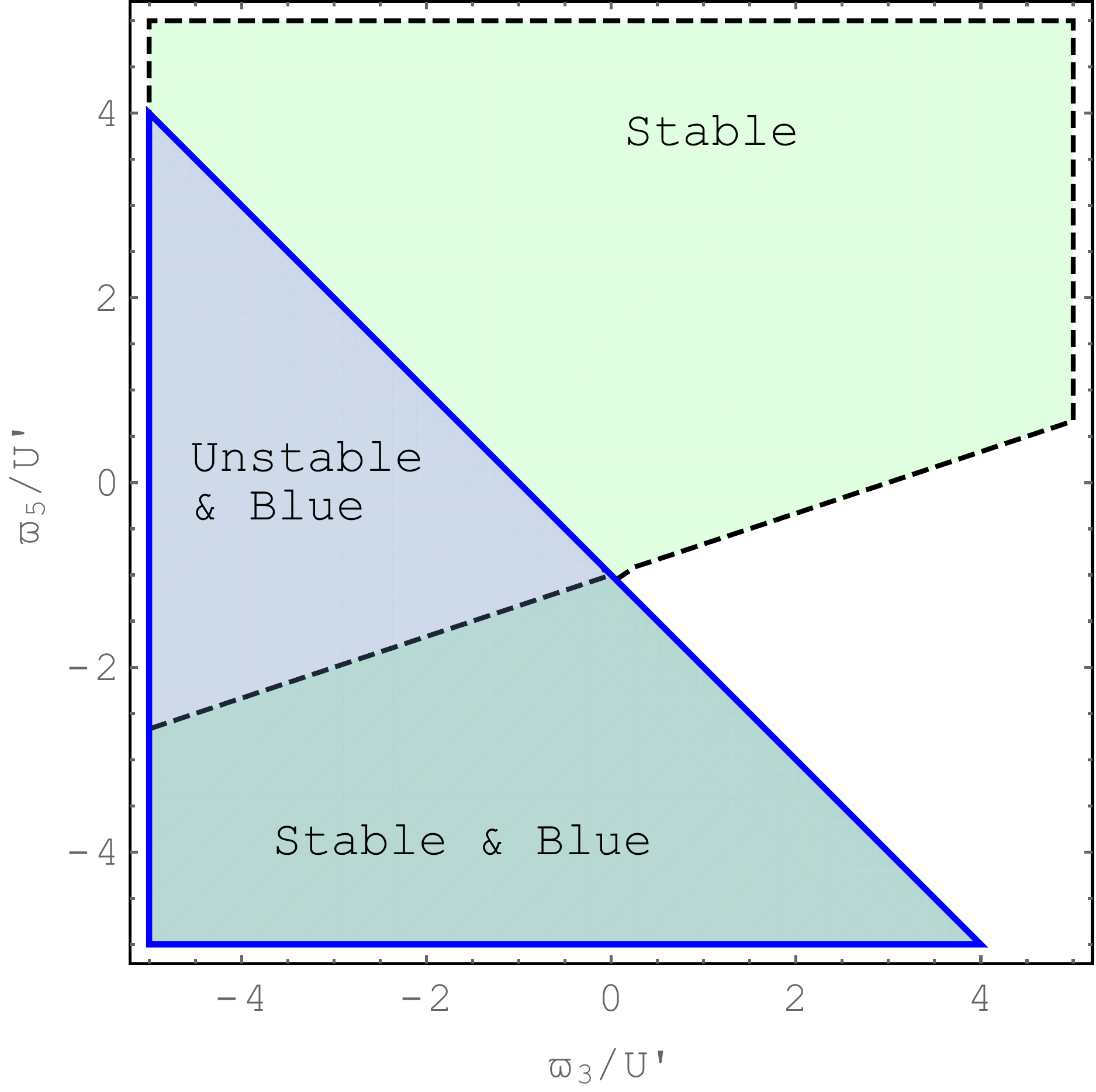}
    \end{center}
       \caption{Tensor tilt and stability in $v=0$ models.
  	}
       \label{fig:fig-lnx.pdf}
  \end{figure}

In this case, one can easily see how
the $\ln X$ terms help to realize $n_t>0$.
The stability conditions read
\begin{align}
u>0 \quad\textrm{and}\quad  \frac{\varpi_3}{U'+\varpi_3+\varpi_5}<\frac{3}{4}.
\label{cond_st}
\end{align}
From Eq.~\eqref{iv0} and $u>0$ we see that
the tensor tilt depends on the sign of $U'/(U'+\varpi)$.
We have a blue tensor spectrum if
\begin{align}
\frac{U'}{U'+\varpi_3+\varpi_5}<0.
\label{cond_bl}
\end{align}
Equations~\eqref{cond_st} and~\eqref{cond_bl}
are satisfied simultaneously if
\begin{align}
\frac{\varpi_3}{U'}-3\frac{\varpi_5}{U'}>3
\quad \textrm{and}\quad
\frac{\varpi_3}{U'}+\frac{\varpi_5}{U'}<-1.\label{condv0inf}
\end{align}
Figure~\ref{fig:fig-lnx.pdf} summarizes the
stability and the tensor tilt in the $(\varpi_3/U',\varpi_5/U')$ plane.
Note that from the definitions and Eqs.~\eqref{sceq1} and~\eqref{2deIII}
we see that $\varpi_3/U'=-3\delta_3/(2\epsilon+\delta_M)$
and $\varpi_5/U'=-2\delta_5/(2\epsilon+\delta_M)$,
and hence we typically have $|\varpi_3/U'|,|\varpi_5/U'|\lesssim {\cal O}(1)$.

\vspace{8mm}

Let us finally check that the result of Gauss-Bonnet inflation
studied in~\cite{Satoh:2008ck,Satoh:2010ep,Koh:2018qcy} can be reproduced from
our general framework.
The Lagrangian considered in~\cite{Koh:2018qcy} is given by\footnote{In terms of the notation of Koh \textit{et al.}~\cite{Koh:2018qcy}, $f(\phi)\to-\xi(\phi)/2$.
Note also that $\epsilon_\textrm{Koh}=\epsilon$,
$\eta_\textrm{Koh}=-2\epsilon + \dot\epsilon/H\epsilon$,
$\delta_{1\textrm{Koh}}=\delta_5$, $\delta_{2\textrm{Koh}}=\epsilon+\dot\delta_5/H\delta_5$.}
\begin{align}
{\cal L}&=\frac{\mpl^2}{2}R+X-V(\phi)
\notag \\ &\quad
+f(\phi)\left(
R^2-4R_{\mu\nu}R^{\mu\nu}+R_{\mu\nu\rho\sigma}R^{\mu\nu\rho\sigma}
\right).
\end{align}
Ignoring the higher-order terms in the slow-roll approximation,
this corresponds to
\begin{align}
&G_2=-V+X,\quad G_3=0,\quad G_4=\frac{\mpl^2}{2},
\notag \\ &
   G_5=-4f'\ln X,
\end{align}
and hence
\begin{align}
u=1,\quad v=0, \quad U'=V', \quad \varpi=\varpi_5=-\frac{8}{3}\frac{V^2f'}{\mpl^4}.
\end{align}
It follows from Eq.~\eqref{condv0inf} that
stable inflation with $n_t>0$ is realized if
\begin{align}
\frac{\varpi_5}{U'}<-1\;\;\;\Leftrightarrow\;\;\;\frac{f'}{\mpl^4}
\frac{V^2}{V'} > \frac{3}{8}.
\end{align}
This reproduces the result of~\cite{Koh:2018qcy}.

The expression for the scalar spectral index can also be reproduced.
Using Eq.~\eqref{eveq}, we obtain $\dot\phi^2=\mpl^2H^2(2\epsilon-\delta_5)$,
and hence ${\cal F}=\mpl^2(\epsilon-\delta_5/2)$.
This, together with Eq.~\eqref{nsformula}, leads to
\begin{align}
n_s-1=-2\epsilon
-\frac{(2\epsilon-\delta_5)\dot{}}{(2\epsilon -\delta_5)H},
\end{align}
which agrees with the result of~\cite{Koh:2018qcy}.
In our notation, this can also be written in a simpler way as
$n_s-1=-4\epsilon +2\eta$.
(See Ref.~\cite{Wu:2017joj} for
a higher-order calculation of the power spectra
in Gauss-Bonnet inflation.)

We have thus successfully extended the work of~\cite{Kamada:2012se}
to include Gauss-Bonnet inflation as a specific case.
It should be emphasized that
our framework not only contains Gauss-Bonnet inflation but also
covers a model space
of potential-driven slow-roll inflation
which has not been explored before.

\section{Conclusions}\label{Section:V}

In this paper, we have explored the possibility of
generating primordial gravitational waves with
a positive tilt, $n_t>0$, from potential-driven slow-roll inflation.
In a canonical inflationary setup this is obviously impossible.
While a blue-tilted gravitational wave spectrum can certainly
be obtained in kinetically driven G-inflation~\cite{Kobayashi:2010cm}
and other more or less radical
models~\cite{Gruzinov:2004ty,Endlich:2012pz,Cannone:2014uqa,Bartolo:2015qvr,%
Ricciardone:2016lym,Fujita:2018ehq,Ashoorioon:2014nta},
it has been claimed in~\cite{Kamada:2012se} that
one always has $n_t<0$ in generic potential-driven slow-roll inflation
constructed within the Horndeski
framework~\cite{Horndeski:1974wa,Deffayet:2011gz,Kobayashi:2011nu}.
It was pointed out, however, that the tensor tilt can
indeed be positive in potential-driven slow-roll inflation
with the Gauss-Bonnet term~\cite{Satoh:2008ck,Satoh:2010ep,Koh:2018qcy}.
Since the nonminimal coupling to the Gauss-Bonnet term
is a (nontrivial) specific case of the Horndeski Lagrangian,
this result implies that
the validity of the statement of~\cite{Kamada:2012se}
is questionable. In this work, we have therefore extended the
formulation of~\cite{Kamada:2012se} in two ways.
First, the Taylor-expanded form of the functions
in the Horndeski Lagrangian assumed in~\cite{Kamada:2012se}
fails to reproduce inflation with the Gauss-Bonnet term
and so we have included new terms that
are still allowed within the Horndeski framework
and help to recover the Gauss-Bonnet term as a specific case.
Second, we have reconsidered the validity of the inequality
assumed in~\cite{Kamada:2012se} and found that
it is in fact too strong.
We have shown that
a blue gravitational wave spectrum can be obtained
even in potential-dominated slow-roll inflation
if one relaxes at least \emph{either} of these two assumptions.
We have thus enlarged a possible model space
of slow-roll inflation with observationally interesting predictions.

\acknowledgments
We thank Filippo Vernizzi for fruitful discussions.
The work of TK was supported by
MEXT KAKENHI Grant Nos.~JP15H05888, JP17H06359, JP16K17707, and JP18H04355.


\appendix
\section{The Einstein frame} \label{Appendix:A}

In this appendix, we discuss the frame (in)dependence of
our main result.

One can move to the Einstein frame for the tensor modes
by performing a conformal transformation,
\begin{align}
g_{\mu\nu}\to \tilde g_{\mu\nu}=\Omega^2(\phi) g_{\mu\nu},
\quad \Omega = \frac{\sqrt{2g_4}}{\mpl}.\label{def:conft}
\end{align}
The time coordinate and the scale factor in the Einstein frame is given by
\begin{align}
\D \tilde t = \Omega(\phi(t))\D t,\quad \tilde a(\tilde t) = \Omega(\phi(t))a(t).
\end{align}
The Hubble parameter in the Einstein frame is obtained as
\begin{align}
\tilde H = \frac{\D\ln \tilde a}{\D \tilde t}=\frac{H}{\Omega}
\left(1+\frac{\delta_M}{2}\right)\simeq \frac{H}{\Omega}.\label{def:tilH}
\end{align}
In terms of the tilde variables, the quadratic action
for the tensor perturbations~\eqref{tensor-action}
is expressed as
\begin{align}
  \tilde S^{(2)}_h=\frac{\mpl^2}{8}\int\D \tilde t\D^3x\, \tilde a^3
  \left[
(  \partial_{\tilde t}h_{ij})^2 - \tilde a^{-2}(\partial_kh_{ij})^2
  \right],\label{tensor-action-Einstein}
\end{align}
and hence is of the standard form~\cite{Creminelli:2014wna}.
The power spectrum in the Einstein frame is thus of the standard form,
\begin{align}
\tilde {\cal P}_h=\frac{2}{\pi^2}\frac{\tilde H^2}{\mpl^2}.
\end{align}
From Eqs.~\eqref{def:conft} and~\eqref{def:tilH},
we see that this coincides with the Jordan frame result~\eqref{eq:power_in_J}.
Obviously, the spectral index in the Einstein frame
coincides with that in the Jordan frame,
\begin{align}
\tilde n_t=-2\tilde\epsilon \simeq -2\epsilon-\delta_M = n_t.
\end{align}
This in particular means that,
in order to have $\tilde n_t>0$,
the null energy condition must be violated in the Einstein frame~\cite{Creminelli:2014wna}.
Note, however, that the conformal transformation~\eqref{def:conft}
preserves the essential nonstandard structure of the scalar sector
rather than leads to its standard form.
Therefore, one may still have stable inflation models with $\tilde n_t>0$.

To see this, let us define
\begin{align}
\tilde V=\frac{V}{\Omega^4},\quad \tilde u=\Omega^2u,\quad \tilde v=\Omega^6 v,
 \quad \tilde\varpi_{3,5} =\frac{\varpi_{3,5}}{\Omega^2}.
\end{align}
and make the field redefinition via
\begin{align}
\D \varphi = \frac{\D\phi}{\Omega^2}.
\end{align}
The background equations in terms of the tilde variables read
\begin{align}
3\mpl^2 \tilde H^2&\simeq \tilde V,
\\
-2\mpl^2\dot{\tilde H}&\simeq \dot\varphi^2\tilde{\cal I},
\\
3\tilde H\dot\varphi\tilde{\cal I}&\simeq-\frac{\D\tilde V}{\D\varphi},
\end{align}
where
\begin{align}
\tilde {\cal I} := \tilde u+3\tilde v\tilde H  \dot \varphi
+\frac{\tilde\varpi_3+\tilde\varpi_5}{3\tilde H\dot\varphi}=\Omega^2{\cal I},
\end{align}
and a dot now stands for differentiation with respect to $\tilde t$.
The information of the nonstandard background evolution
is encoded in $\tilde{\cal I}$.
Similarly, the quadratic action for the curvature perturbation
in terms of the tilde variables is given by
\begin{align}
\tilde S_\zeta^{(2)}=\int \D \tilde t\D^3x\, \tilde a^3\frac{\tilde {\cal F}}{\tilde c_s^2}
\left[\dot\zeta^2-\tilde a^{-2}\tilde c_s^2(\partial\zeta)^2\right],
\label{scalar-action-ein}
\end{align}
where
\begin{align}
\tilde {\cal F}&=\frac{\dot\varphi^2}{2\tilde H^2}\left(\tilde u+
4\tilde v\tilde H\dot\varphi +\frac{4\tilde\varpi_3}{9\tilde H\dot\varphi}\right)
=\frac{{\cal F}}{\Omega^2},
\\
\tilde c_s^2&=
\frac{\tilde u+4\tilde v\tilde H\dot\varphi
 + 4\tilde\varpi_3/9\tilde H\dot\varphi}{\tilde u+6\tilde v\tilde H\dot\varphi}
 =c_s^2.
\end{align}
With Eq.~\eqref{def:tilH},
this clearly shows that the stability and observables are
frame-independent.

\bibliography{refs}
\bibliographystyle{JHEP.bst}
\end{document}